\documentstyle[11pt]{article}

\makeatletter
\@addtoreset{equation}{section}
\makeatother

\def\dalemb#1#2{{\vbox{\hrule height .#2pt
        \hbox{\vrule width.#2pt height#1pt \kern#1pt
                \vrule width.#2pt}
        \hrule height.#2pt}}}

\let\a=\alpha

\def\nn{\nonumber} \def\bd{\begin{document}} \def\ed{\end{document}}
\def\ds{\documentstyle} \let\fr=\frac \let\bl=\bigl \let\br=\bigr
\let\Br=\Bigr \let\Bl=\Bigl
\let\bm=\bibitem
\let\na=\nabla
\let\pa=\partial \let\ov=\overline
\newcommand{\be}{\begin{equation}}
\newcommand{\ee}{\end{equation}}
\def\ba{\begin{array}}
\def\ea{\end{array}}
\def\ft#1#2{{\textstyle{{\scriptstyle #1}\over {\scriptstyle #2}}}}
\def\fft#1#2{{#1 \over #2}}
\def\del{\partial}
\def\ie{{\it i.e.\ }}
\def\sst#1{{\scriptscriptstyle #1}}
\def\oneone{\rlap 1\mkern4mu{\rm l}}
\def\Z{\rlap{\ssr Z}\mkern3mu\hbox{\ssr Z}}
\def\R{\rlap{\rm I}\mkern3mu{\rm R}}
\newcommand{\ho}[1]{$\, ^{#1}$}
\newcommand{\hoch}[1]{$\, ^{#1}$}
\newcommand{\bea}{\begin{eqnarray}}
\newcommand{\eea}{\end{eqnarray}}
\newcommand{\ra}{\rightarrow}
\newcommand{\lra}{\longrightarrow}
\newcommand{\Lra}{\Leftrightarrow}
\newcommand{\ap}{\alpha^\prime}
\newcommand{\bp}{\tilde \beta^\prime}
\newcommand{\tr}{{\rm tr} }
\newcommand{\Tr}{{\rm Tr} }
\newcommand{\NP}{Nucl. Phys. }
\newcommand{\tamphys}{\it Center for Theoretical Physics,
Texas A\&M University, College Station, Texas 77843}

\newcommand{\auth}{I.V. Lavrinenko, H. L\"u{\hoch{\dagger}} and 
C.N. Pope{\hoch{\dagger}}}

\begin{document}

\hfill{CTP-TAMU-57/96}

\hfill{hep-th/9611134}

\vspace{20pt}

\begin{center}
{ \large {\bf From Topology to Generalised Dimensional Reduction}}

\vspace{30pt}
\auth

\vspace{15pt}
{\tamphys}
\vspace{40pt}

\underline{ABSTRACT}
\end{center}
\vspace{40pt}

     In the usual procedure for toroidal Kaluza-Klein reduction, all the 
higher-dimensional fields are taken to be independent of the coordinates on 
the internal space.  It has recently been observed that a generalisation of 
this procedure is possible, which gives rise to lower-dimensional {\it 
massive} supergravities.  The generalised reduction involves allowing  
gauge potentials in the higher dimension to have an additional linear 
dependence on the toroidal coordinates.  In this paper, we show that a much 
wider class of generalised reductions is possible, in which 
higher-dimensional potentials have additional terms involving differential 
forms on the internal manifold whose exterior derivatives yield 
representatives of certain of its cohomology classes.  We consider various 
examples, including the generalised reduction of M-theory and type II 
strings on K3, Calabi-Yau and 7-dimensional Joyce manifolds.  The resulting 
massive supergravities support domain-wall solutions that arise by the vertical 
dimensional reduction of higher-dimensional solitonic $p$-branes and 
intersecting $p$-branes.

{\vfill\leftline{}\vfill
\vskip	10pt
\footnoterule
{\footnotesize	\hoch{\dagger} Research supported in part by DOE
Grant DE-FG05-91-ER40633 \vskip	-12pt}}

\pagebreak
\setcounter{page}{1}

\section{Introduction}

        M-theory or string theory are generally believed to be the
fundamental theories that may provide a basis for quantising Einstein's
general relativity.   One of the reasons for the recent upsurge of interest
in the subject has been the discovery of duality relations between string
theories that were at one time believed to be independent. One of the ways
in which these duality relations can be brought to light is by studying the
dimensionally reduced theories that are obtained by compactifying on certain
internal spaces.  The kind of dimensional reduction that is usually
considered is the standard Kaluza-Klein procedure, in which the fields of
the higher-dimensional theory are first expanded in terms of complete sets
of harmonics on the internal space, followed by a truncation to the massless
sector of the resulting lower-dimensional theory.   It is crucial that this
truncation be {\it consistent}, which means that all solutions of the
lower-dimensional truncated theory should also be solutions of the
higher-dimensional one. Only then will the properties of the
lower-dimensional theory reflect themselves in corresponding properties of
the higher-dimensional theory. The criterion for consistency is that when
one considers the equations of motion for the lower-dimensional massive
fields prior to truncation, there should be no source terms constructed
purely from the massless fields that are to be retained.  This consistency
of the truncation is obvious for the case of toroidal compactifications with
the standard Kaluza-Klein ansatz, where the higher-dimensional fields are
simply taken to be independent of the toroidal coordinates.  In more
complicated cases such as compactifications on K3 or Calabi-Yau spaces,
consistency is not so obvious, since products of the zero-mode harmonics can
generate non-zero-mode harmonics.  It appears that in fact supersymmetry
plays a crucial in establishing the consistency of the truncation in these
cases \cite{ps,dfps}. 

     It was recently observed that the usual Kaluza-Klein ansatz for
toroidal compactification is slightly more restrictive than is actually
necessary in order to achieve consistency of the truncation. Specifically,
it was shown in \cite{bdgpt} that in the dimensional reduction of the type
IIB theory on a circle, one can allow the Ramond-Ramond axion $\chi$ to have
an additional linear dependence on the coordinate $z$ of the circle,
$\chi(x,z) \rightarrow m\, z + \chi(x)$, where $x$ denotes the
lower-dimensional coordinates and $m$ is a constant parameter.  This does
not upset the consistency of the reduction, since $\chi$ appears in the
ten-dimensional equations of motion only through its derivative $d\chi$, and
thus there is still no $z$-dependence, even when $m$ is non-zero.  The
resulting nine-dimensional theory is a massive supergravity, with a
cosmological term \cite{bdgpt}.   In fact, it is T-dual to the theory that
one obtains by performing a standard Kaluza-Klein reduction of the massive
IIA supergravity \cite{r} on a circle.  This generalisation of the
Kaluza-Klein ansatz for compactification on a circle was subsequently
applied to the dimensional reduction of all other $D\le 11$ maximal
supergravities in \cite{clpst}. (In fact generalised Kaluza-Klein ans\"atze
that give rise to cosmological terms were also discussed in a general
group-theoretic framework in \cite{ss}.  It was also observed in
\cite{dk,dkmr}, in the context of compactifying the heterotic string to $D=4$,
that wrapping the 5-brane on a 3-torus to give rise to a membrane in $D=4$
would require some ansatz that went beyond the usual Kaluza-Klein
dimensional reduction.) 

     Consistent generalisations of the Kaluza-Klein ansatz are possible in 
more complicated compactifications also.  In \cite{lpdomain}, it was shown
that the generalised ansatz for the axion $\chi$ discussed above is a
special case of the ansatz $A_{n-1}(x,z) = m\, \omega_{n-1}(z) + {\rm
standard\  terms}$, where $A_{n-1}$ is an $(n-1)$-form potential in the
higher-dimensional theory, which is then compactified on an $n$-dimensional
internal manifold $M_n$ with coordinates $z$, whose volume form $\Omega_n$
is given locally by $d\omega_{n-1}$.  (In the case $M_1=S^1$, we can
represent $\omega_0$ locally by $\omega_0=z$, giving the globally-defined
volume form $\Omega_1=dz$.)  Again, provided that $A_{n-1}$ appears always
through its exterior derivative $dA_{n-1}$ in the higher-dimensional
equations of motion, these equations will depend only on the zero-mode
harmonics on $M_n$ after imposing the generalised ansatz, and thus the
truncation will still be consistent. 

     In all cases, the generalised dimensional reduction procedure gives
rise to a massive supergravity with a cosmological term, of the form $-\ft12 
m^2\, e\, e^{a\phi}$, where $\phi$ is some dilatonic scalar field.  Such 
theories admit no maximally-symmetric Minkowski or anti-de Sitter ground 
state.  Instead, the most symmetrical, and therefore most natural, ground 
state solution is a domain wall, which is a $(D-2)$-brane in $D$ dimensions.  
In fact this domain wall solution arises from the vertical dimensional 
reduction of a standard kind of $p$-brane soliton in the higher dimension.
The field strength that supports this higher-dimensional $p$-brane solution has 
a form that is compatible with the generalised Kaluza-Klein ansatz for its 
potential.  Indeed, it is compatible {\it only} with the generalised ansatz, 
and in fact the original motivation for considering such a generalisation 
was in order to explain how a 7-brane in the type IIB string could be 
vertically reduced to a solution of a theory in $D=9$ \cite{bdgpt}.

     In this paper, we shall develop these ideas further, by showing that 
one can make use of representatives of some appropriate cohomology classes 
of the compactifying manifold $M_n$, enlarging considerably the number of 
possibilities for generalised Kaluza-Klein reductions.  In section 2, we 
shall explain the basic idea, and apply it to compactifications of the type 
IIA string.  Then, in section 3, we shall consider applications to the 
compactification of M-theory, followed by compactifications of the type IIB
theory in section 4.  In section 5, we end the paper with a discussion of
the supersymmetry of the compactified theories, and duality. 

\section{Generalised dimensional reduction of type IIA strings}

     We begin our discussion by considering the case of the type IIA
string, and its generalised dimensional reduction on certain Ricci-flat
compact manifolds $M_n$.  In particular, we shall consider its
compactification on $T^4$, K3 and six-dimensional Calabi-Yau spaces. The
low-energy limit of the type IIA string is type IIA supergravity, whose
bosonic sector contains the metric and dilaton, together with a rank 4, a
rank 3 and a rank 2 field strength.  We shall denote these by
$g_{\sst{MN}}$, $\phi_1$, $F_4$, $F^{(1)}_3$ and ${\cal F}_2^{(1)}$
respectively.  This notation is derived from the fact that type IIA
supergravity itself can be obtained by dimensional reduction from $D=11$
supergravity.  Using the notation introduced in \cite{lpsol}, the toroidal
reduction from $D=11$ to $D$ dimensions gives field strengths $F_4$,
$F_3^{(i)}$, $F_2^{(ij)}$, $F_1^{(ijk)}$ from the 4-form $F_4$ in $D=11$,
and field strengths ${\cal F}_2^{(i)}$ and ${\cal F}_1^{(ij)}$ from the
$D=11$ metric.  In addition, there will be dilatonic scalars $\phi_i$.  The
index $i$ runs over the $11-D$ toroidally-compactified coordinates $z_i$. 

     Using the above notation, the bosonic sector of type IIA supergravity 
becomes
\bea
\hat e^{-1} \hat {\cal L} &=& \hat R -\ft12 (\del\phi_1)^2 -\ft1{48}
e^{-\fft12 \phi_1} \, \hat F_4^2 -\ft1{12} e^{\phi_1}\, (\hat F^{(1)}_3)^2
-\ft14 e^{-\ft32\phi_1} \, (\hat{\cal F}^{(1)})^2 \label{d10lag}\\ &&+\ft12
\hat e^{-1}\, d\hat A_3\wedge d\hat A_3 \wedge \hat A_2^{(1)}\ , 
\eea
where the final term is presented as a 10-form.  The field strengths are 
given by
\be
\hat F_4=d\hat A_3 + \hat A_2^{(1)}\wedge d \hat {\cal A}_1^{(1)}\ ,
\qquad \hat F^{(1)}_3 
=d\hat A_2^{(1)}\ ,\qquad \hat{\cal F}_2^{(1)} = d\hat{\cal A}_1^{(1)}\ .
\label{csd10}
\ee

\subsection{$T^4$ and K3 compactifications}

    The standard $T^4$ compactification, using the usual Kaluza-Klein 
ansatz, gives rise to maximal $N=4$ supergravity in $D=6$.  Various 
generalisations of this reduction are possible, in which a 
higher-dimensional gauge potential acquires an additional linear dependence 
on one or more of the coordinates of the 4-torus.  These generalised 
reductions give rise to massive supergravities in $D=6$, with cosmological 
terms.  One example that has recently been discussed involves taking the 
ansatz for the 3-form potential in $D=10$ to have the generalised form
\be
A_3^{(1)}(x,z)=m\, \omega_3 + A_3^{(1)}(x) +\cdots\ ,\label{4torus}
\ee
where $d\omega_3=\Omega_4=dz_2\wedge dz_3\wedge dz_4\wedge dz_5$ is the 
volume form on the 4-torus \cite{lpdomain}.  One may choose to write
$\omega_3$ locally as, for example, $\omega_3 = z_2\, dz_3\wedge dz_4\wedge
dz_5$. All the other ten-dimensional fields are reduced using the standard
Kaluza-Klein ansatz, in which there is no dependence on any of the $z_i$
coordinates.  This generalised dimensional reduction gives rise to a massive
maximal supergravity in $D=6$ with a cosmological term.  Its natural ground
state is a domain wall solution, which is the vertical dimensional reduction
of the solitonic 4-brane in $D=10$.  Note that this massive supergravity
theory does not have any ground state that admits the full set of 32 Killing
spinors that arise in the Minkowski ground state of massless $N=4$
supergravity in $D=6$.  (Nor does it admit an anti-de Sitter ground state,
since the cosmological term has a dilaton coupling.) The domain wall ground
state admits 16 Killing spinors.  Even though there is no
maximally-symmetric ground state in the massive theory, we shall
nevertheless adopt the standard practice of describing this solution as
preserving one half of the global supersymmetry. 

   The generalised reduction discussed above uses the cohomology
class $H^4(M_4,\R)$, representing the fact that the volume form $\Omega_4$
has a non-zero integral over the entire 4-manifold.  In this section, we
consider another type of ansatz for the generalised reduction, which
makes use of the cohomology class $H^2(M_4,\R)$.  We shall apply this idea
in two examples, beginning with the 4-torus, and then K3. The generalised
ansatz will now be made for the 1-form potential rather than the
3-form.\footnote{Since the 4-torus also has non-vanishing $H^3$ cohomology,
we could also consider generalising the ansatz for the 2-form potential in
$D=10$.  We shall not pursue this further, since it does not extend to the
K3 compactification.}   In the case of the torus, there are six independent
2-forms $dz_i\wedge dz_j$ in $H^2(T^4,\R)$, where $2\le i<j\le 5$, allowing
a six-parameter family of generalised ans\"atze for the 1-form potential.
We shall consider the following two-parameter example, 
\be
{\cal A}^{(1)}_1(x,z) = m_1\, z_2\, dz_3 + m_2\, z_4\, dz_5 + 
{\cal A}^{(1)}_1(x) + {\cal A}_0^{(1j)}\, dz_j \ ,\label{a1ans}
\ee
where the summation in the final term is over $2\le j \le 5$.  All the other 
ten-dimensional fields are reduced using the usual $z$-independent 
ans\"atze,
\bea
A_3(x,z) &=& A_3(x) + A_2^{(i)}(x)\wedge dz_i +\ft12 A_1^{(ij)}\wedge 
dz_i \wedge dz_j +\ft16 A_0^{(ijk)}(x)\, dz_i\wedge dz_j\wedge dz_k \ ,\nn\\
A_2^{(1)}(x,z) &=& A_2^{(1)}(x) + A_1^{(1j)}(x)\wedge dz_j + \ft12 
A_0^{(1jk)}(x) \, dz_j\wedge dz_k\ ,\label{a32ans}
\eea
where the indices run over the range 2 to 5.  The metric ansatz is
\bea
ds_{10}^2 &=& e^{2\a_2\phi_2 +\cdots 2\a_5\phi_5} \, ds_6^2 + e^{2\a_3\phi_3 
+\cdots 2\a_5\phi_5 -14\a_2\phi_2}\, h_5^2 \ ,\nn\\
&&+e^{2\a_4\phi_4 +2\a_5\phi_5 -12\a_3\phi_3}\, h_4^2 +
e^{2\a_5\phi_5 -10\a_4\phi_4}\, h_3^2 + e^{-8\a_5\phi_5}\, h_2^2\ ,
\label{metricans}
\eea
where $(\a_i)^{-2} = 2 (10-i)(9-i)$ and the scalars $\phi_i$ and the metric 
$ds_6^2$ are functions only of the $x$ coordinates.  The 1-forms $h_i$ are
given by 
\be 
h_i=dz_i + {\cal A}_1^{(i)}(x) + {\cal A}_0^{(ij)}(x)\, dz_j\ ,\label{hdef}
\ee
where the summation is over $i<j\le 5$, since ${\cal A}_0^{(ij)}$ is defined 
only for $i<j$.  This can easily be inverted to 
express $dz_i$ in terms of $h_i$ \cite{lpsol}:
\be
dz_i=\gamma_{ij}\, (h_j -{\cal A}_1^{(j)}) \ ,\label{zsol}
\ee
where the summation is over $i\le j\le 5$, and 
\be
\gamma_{ij}=[(1+{\cal A}_0)^{-1}]^{ij}=\delta^{ij} -{\cal A}_0^{(ij)} +
{\cal A}_0^{(ik)} {\cal A}_0^{(kj)} + \cdots\ .\label{gamma}
\ee

     Substituting all the above ans\"atze into the Lagrangian 
(\ref{d10lag}), we obtain a six-dimensional Lagrangian that describes a 
consistent truncation of type IIA supergravity, since it is independent of 
the $z$ coordinates.  The full expression for the six-dimensional Lagrangian 
is quite complicated, and we shall not present it explicitly here.  Instead, 
we shall concentrate on the essential features, in particular describing the
spectrum of massive and massless fields.  When the substitution of the
ans\"atze is performed, one first obtains an expression in $D=6$ with a
kinetic term for each of the field strengths formed from the six-dimensional
potentials.  However, these field strengths will in general have
Chern-Simons modifications.  To see how these arise, it is simplest to
express the ten-dimensional field strengths in terms of a tangent-space
basis, since then the expressions for their kinetic terms appearing in
(\ref{d10lag}) can immediately be read off.  Thus for example, it follows
from (\ref{a32ans}) that the 3-form field strength in $D=10$ becomes 
\be
F_3^{(1)}(x,z) = dA_2^{(1)}(x) + dA_1^{(1j)}(x)\wedge dz_j 
  +\ft12  dA_0^{(1jk)}\, dz_j\wedge dz_k\ , 
\ee
which can then be expressed in the tangent-space basis by replacing $dz_i$
by $h_i$, using (\ref{zsol}).  For the purposes of determining the spectrum
of massive and massless fields in $D=6$, it suffices to look only at the
terms up to linear order in fields, since these will govern the form of the
kinetic terms.  Thus we find that all of the field strengths in $D=6$ are
given by $F=dA+$ 2'nd order terms, with the following exceptions: 
\bea
&&F_2^{(23)}\sim dA_1^{(23)} + m_1\, A_2^{(1)} \ ,\qquad\quad\quad
F_2^{(45)} \sim dA_1^{(45)} + m_2\, A_2^{(1)} \ ,\nn\\
&&F_1^{(234)} \sim dA_0^{(234)} + m_1\, A_1^{(14)} \ ,\qquad
F_1^{(235)} \sim dA_0^{(235)} + m_1\, A_1^{(15)} \ ,\nn\\
&&F_1^{(245)} \sim dA_0^{(245)} + m_2\, A_1^{(12)} \ ,\qquad
F_1^{(345)} \sim dA_0^{(345)} + m_2\, A_1^{(13)} \ ,\label{lin}\\
&&{\cal F}_1^{(12)} \sim d{\cal A}_0^{(12)} + m_1\, {\cal A}_1^{(3)}\ ,
\qquad\quad\quad {\cal F}_1^{(13)} \sim d{\cal A}_0^{(13)} - 
m_1\, {\cal A}_1^{(2)}\ ,\nn\\
&&{\cal F}_1^{(14)} \sim d{\cal A}_0^{(14)} + m_2\, {\cal A}_1^{(5)}\ ,
\qquad\quad\quad {\cal F}_1^{(15)} \sim d{\cal A}_0^{(15)} + m_2\, 
{\cal A}_1^{(4)}\ .\nn
\eea
Here, the $\sim$ symbol indicates that we have omitted the terms of 2'nd 
order and above.

     We can see from (\ref{lin}) that, at least to the leading order, the 
potentials appearing through their exterior derivatives on the right-hand 
sides can be absorbed by making gauge transformations of the associated
undifferentiated potentials.  After doing this, the previous kinetic terms
for the fields that are absorbed will become mass terms for the potentials
that absorb them.  In other words, the former fields are eaten by the
latter, in order for them to become massive.  Consistency of the theory
requires that the fields that are eaten should then disappear everywhere
from the Lagrangian. To clarify this phenomenon, and to show that it works
to all orders in the various fields, we may exhibit explicitly the relevant
local symmetries of the Lagrangian in $D=6$ that can be used to eliminate
the fields.  These symmetries originate from gauge symmetries of the
ten-dimensional theory.  After the generalised dimensional reduction, they
become St\"uckelburg symmetries in $D=6$, under which the fields that are
eaten undergo pure non-derivative shift symmetries.  The symmetries can
therefore be used to set these fields to zero. 

     The relevant local symmetries in $D=10$ are included among the gauge
transformations of the potentials $\hat A_3$, $\hat A_2^{(1)}$ and 
$\hat{\cal A}_1^{(1)}$, and certain general coordinate transformations of the
compactified coordinates: 
\bea
\hat A_3&\rightarrow &\hat A_3' =\hat A_3 + d\hat\Lambda_2 - 
\hat\Lambda_1^{(1)} \wedge d\hat {\cal A}_1^{(1)} \ ,\nn\\
\hat A_2^{(1)} &\rightarrow &\hat {A_2^{(1)}}' = \hat A_2^{(1)} + d \hat 
\Lambda_1^{(1)}\ ,\nn\\
\hat {\cal A}_1^{(1)} &\rightarrow &\hat {{\cal A}_1^{(1)}}'= \hat{\cal 
A}_1^{(1)} + d\hat\Lambda_0^{(1)} \ ,\label{gauge}\\
z_i &\rightarrow & z_i' =z_i + \xi_i(x)\ .
\eea
The gauge parameters are dimensionally reduced in the same way as the gauge 
potentials, namely $\hat\Lambda_1^{(1)} = \Lambda_1^{(1)}(x) +
\Lambda_0^{(1j)}(x) dz_j$, {\it etc}. 

     Consider first the lower-dimensional gauge parameters
$\Lambda_1^{(1)}(x)$.  These generate the transformations 
\bea 
&&{A_1^{(23)}}'= A_1^{(23)} - m_1\, \Lambda_1^{(1)} \ ,\qquad
{A_1^{(45)}}'= A_1^{(45)} - m_2\, \Lambda_1^{(1)} \ ,\nonumber\\
&& A_3' = A_3 - \Lambda_1^{(1)} \wedge d{\cal A}_1^{(1)}\ ,\qquad
{A_2^{(1)}}' = A_2^{(1)} + d\Lambda_1^{(1)}\ ,\label{gauge1}\\
&&{A_2^{(i)}}' = A_2^{(i)} -\Lambda_1^{(1)}\wedge d{\cal A}_0^{(1i)}\ ,
\qquad i=2,\ldots, 5\ ,\nonumber
\eea
where we have displayed only those fields on which the transformation acts
non-trivially. We see that the ten-dimensional gauge symmetry associated
with the gauge parameter $\Lambda_1^{(1)}$ becomes a St\"uckelburg symmetry
for the potentials $A_{1}^{(23)}$ and $A_{1}^{(45)}$.  Thus we can choose
the parameter $\Lambda_1^{(1)}$ appropriately so as to set either
$A_1^{(23)}$ or $A_1^{(45)}$, but not both, to zero.  Either way, it has the
consequence that the potential $A_2^{(1)}$ becomes massive. 

    Analogously, the low-dimensional gauge parameters $\Lambda_0^{(1i)}(x)$, 
for $i=2,\ldots, 5$, also generate St\"uckelburg symmetries:
\bea
&&{A_0^{(i23)}}' = A_0^{(i23)} - m_1 \Lambda_0^{(1i)}\ ,\quad
i=4,5\ ,\qquad
{A_0^{(i45)}}' = A_0^{(i45)} - m_2 \Lambda_0^{(1i)}\ ,\quad
i=2,3\ ,\nonumber\\
&&{A_1^{(ij)}}' = A_1^{(ij)} + \Lambda_0^{(1i)} d{\cal A}_0^{(1j)} -
\Lambda_0^{(1j)} d{\cal A}_0^{(1i)}\ ,\qquad i,j=2,\ldots, 5\label{gauge2}\\
&&{A_2^{(i)}} = A_2^{(i)} - \Lambda_0^{(1i)} d{\cal A}_1^{(1)}\ ,
\qquad i=2,\ldots, 5\nonumber
\eea
Thus we can set the potentials $A_0^{(234)}$, $A_0^{(235)}$, $A_0^{(245)}$
and $A_0^{(345)}$ to zero and correspondingly the fields $A_1^{(1i)}$ become 
massive, with masses $m_1$ for $i=4,5$ and $m_2$ for $i=2,3$.

     The St\"uckelburg symmetry that can be used to eliminate the potentials
${\cal A}_0^{1i}$ comes from the general coordinate transformations
$z_i' =z_i + \xi_i(x)$ in $D=10$.  Since these leave the left-hand sides of 
the ten-dimensional expansions (\ref{a1ans}), (\ref{a32ans}) and 
(\ref{hdef}) invariant, we can read off their action on the six-dimensional 
fields:
\bea
&&{{\cal A}_0^{(12)}} ={\cal A}_0^{(12)} + m_1 \xi_3\ ,\qquad
{{\cal A}_0^{(13)}}'={\cal A}_0^{(13)} -m_1 \xi_2\ ,\nn\\
&& {{\cal A}_0^{(14)}}'={\cal A}_0^{(12)} + m_2 \xi_5\ ,\qquad
{{\cal A}_0^{(15)}}' ={\cal A}_0^{(15)} - m_1 \xi_4\ ,\nn\\
&&{A_1^{(ij)}}' = A_1^{(ij)} - A_0^{(ijk)} d\xi_k\ ,\qquad
{{\cal A}_1^{(1)}}'={\cal A}_1^{(1)} -{\cal A}_0^{(1i)} d\xi_i
-m_1 \xi_3 d\xi_2 - m_2 \xi_5 d\xi_4\ ,\nn\\
&& {{\cal A}_1^{(i)}}' = {\cal A}_1^{(i)} - d\xi_i
-{\cal A}_0^{(ij)} d\xi_j\ ,\qquad i=2,\ldots, 5\ ,\quad
i\le j \le 5\ ,\\
&&{A_2^{(1)}}' = A_2^{(1)} - A_1^{(1i)} \wedge d\xi_i+
\ft12 A_0^{(1ij)} d\xi_i\wedge d\xi_j\ ,\nn\\
&& {A_2^{(i)}}' = A_2^{(i)} + A_1^{(ij)}\wedge d\xi_i+
\ft12 A_0^{(ijk)} d\xi_j\wedge d\xi_k\ ,\qquad i=2,\ldots, 5\nn\\
&&A_3' = A_3 - A_2^{(i)}\wedge d\xi_i +
\ft12 A_1^{(ij)}\wedge d\xi_i\wedge d\xi_j -
\ft16 A_0^{(ijk)} d\xi_i\wedge d\xi_j\wedge d\xi_k\ .\nn
\eea

     In summary, we have seen that the generalised Kaluza-Klein ansatz 
(\ref{a1ans}) gives masses to the fields $A_2^{(1)}$, $A_1^{(1i)}$ and ${\cal 
A}_1^{(i)}$.  The fields that are eaten in each case can be seen by 
inspection of (\ref{lin}).  In addition, we can see from the expansion of 
(\ref{a1ans}) up to linear order that the four axions ${\cal A}_0^{(24)}$,
${\cal A}_0^{(25)}$, ${\cal A}_0^{(34)}$ and ${\cal A}_0^{(35)}$ acquire 
masses $m_1$.  Since these are scalars, there is no associated St\"uckelburg
symmetry.  There are two cosmological terms in $D=6$, taking the form 
\be 
{\cal L}_{\rm cosmo}= -\ft12 m_1^2\, e\,e^{\vec b_{123}\cdot \vec\phi}
-\ft12 m_2^2\, e\, e^{\vec b_{145}\cdot\vec\phi} \ ,\label{cos} 
\ee
where the dilaton vectors $\vec b_{123}$ and $\vec b_{145}$ are defined in 
\cite{clpst}.

     As we have discussed previously, this massive theory in $D=6$ will 
admit domain wall solutions.  The metric takes the form
\be
ds_6^2 = (H_1\, H_2)^{\ft14}\, dx^\mu\, dx^\nu\, \eta_{\mu\nu} +
( H_1\, H_2)^{\ft54}\, dy^2 \ ,\label{d6dw}
\ee
where $H_1=1+m_1 |y|$ and $H_2=1+m_2 |y|$.  There are two dilatonic scalars, 
$\varphi_1=\vec b_{123}\cdot \vec\phi$ and $\varphi_2=\vec b_{145}\cdot 
\vec\phi$, given by
\be
e^{-\varphi_1} = H_1^{\ft{13}4}\, H_2^{\ft54}\ ,\qquad
e^{-\varphi_2} = H_2^{\ft{13}4}\, H_1^{\ft54}\ .\label{d6dw1}
\ee
The other orthogonal components of $\vec\phi$ are zero.  This solution can be
oxidised back to $D=10$, where the ten-dimensional metric becomes
\bea
ds^2_{10} &=& (H_1 H_2)^{-\ft18} dx^\mu dx^\nu \eta_{\mu\nu} + 
              (H_1H_2)^{\ft78} dy^2\nn\\
&&+           H_1^{\ft78}H_2^{-\ft18} (dz_2^2 + dz_3^2) +
              H_1^{-\ft18} H_2^{\ft78} (dz_4^2 + dz_5^2)\ ,\label{d10sol}
\eea
with the dilaton $\phi_1$ and the 2-form field strength $\hat F_2^{(1)}$
given by 
\be
e^{\phi_1} =(H_1 H_2)^{3/4}\ ,\qquad
\hat F_2^{(1)} = m_1  dz_2\wedge dz_3 + m_2 dz_4\wedge dz_5\ .\label{f2}
\ee
The metric (\ref{d10sol}) describes two intersecting 6-branes in $D=10$.  To
see this, we note that if we set $m_1=0$, so that the harmonic function
$H_1$ equals one, the metric describes 6-branes with world volume
coordinates $(x^\mu, z_2, z_3)$ and transverse space $(y, z_4, z_5)$.  Since
the 6-brane solution is independent of the transverse coordinates $(z_4,
z_5)$, it describes a plane of 6-branes whose charges are uniformly
distributed over $(z_4, z_5)$.  On the other hand, if we instead set
$m_2=0$, and hence $H_2=1$, the solution describes a plane of 6-branes
distributed over $(z_2, z_3)$, with world volume coordinates $(x^\mu, z_4,
z_5)$.   If we now consider the general case where $m_1$ and $m_2$ take
generic non-vanishing values, we see that the solution (\ref{d10sol})
interpolates between these two configurations, and can be interpreted as
describing the intersection of the two six-branes.  Many other examples of
intersecting branes were discussed in [11-16]. 

     So far we have discussed two examples of generalised compactification
of the type IIA theory on a 4-torus.  The first one involves a generalised
ansatz for the 3-form potential, given by (\ref{4torus}), and the second
involves a generalised ansatz for the 1-form potential, given in
(\ref{a1ans}).  In fact we can make both the generalised ans\"atze
simultaneously without spoiling the consistency of the Kaluza-Klein
reduction, since the 3-form potential $\hat A_3$ and the 1-form
potential $\hat {\cal A}_1^{(1)}$ simultaneously appear in the Lagrangian
only through their exterior derivatives.  Furthermore, the ansatz for the
1-form potential given in (\ref{a1ans}) is just an example for the
generalised reduction. The most general ansatz takes the form $\hat {\cal
A}_1^{(1)}= m_{ij}\, z_i\, dz_j + \cdots$. 

     One reason that we have chosen the specific example of the ansatz
(\ref{a1ans}) is that the associated reduced theory in $D=6$ admits a
domain-wall solution, given by (\ref{d6dw}); not all the generalised
reductions of the type IIA theory admit $p$-brane solutions.  Another reason
is that the above example, which is a generalised Kaluza-Klein
compactification on a 4-torus, can easily be extended to a compactification
on the K3 manifold.  To see this we note that in the metric describing two
intersecting 6-branes (\ref{d10sol}), the 4-dimensional compactified space
$(z_2, z_3, z_4, z_5)$ becomes isotropic when $m_1=m_2$,  and the 2-form
field strength given in (\ref{f2}) becomes self-dual with respect to the
four compactified dimensions.   In fact, modulo relabellings of the internal 
coordinates $z_i$, the configuration (\ref{f2}) for $\hat F_2^{(1)}$, with 
two terms spanning the internal 4-dimensional space, is the unique 
possibility that can give the isotropic form for the internal metric.
Thus we will still have a solution if we
replace the compactifying 4-torus with metric $dz_i\, dz_i$ by any compact
4-manifold with Ricci-flat metric $ds_4^2$  which admits a covariantly
constant self-dual 2-form $J$, giving 
\bea
ds_{10}^2 &=& H^{-\ft14}\, dx^\mu\, dx^\nu \, \eta_{\mu\nu} + H^{\ft74} \,
dy^2 + H^{\ft34}\, ds_4^2\ ,\nn\\
e^{\phi_1}&=& H^{\ft34}\ ,\qquad \hat F_2^{(1)} = m\, J\ ,\label{k3sol}
\eea
where $H=1 + m\, |y|$.  In particular, we may take the compactifying space 
to be K3.   Note that when $H_1=H_2=H$, the six-dimensional domain-wall 
solution (\ref{d6dw}) is of the form \cite{lpss1}
\be
ds^2_{6} = H^{\ft{4}{\Delta(D-2)}} dx_\mu dx_\nu \eta_{\mu\nu} +
           H^{\ft{4(D-1)}{\Delta(D-2)}} dy^2\ ,\label{sssol}
\ee
with $D=6$ and $\Delta =2$.  In general, the metrics (\ref{sssol}) are 
domain-wall solutions for the single-scalar Lagrangian 
\be
e^{-1}{\cal L} = R - \ft12 (\del\phi)^2 -\ft12 m^2 e^{a\phi},
\ee
with $a$ parameterised by $a^2 = \Delta + 2(D-1)/(D-2)$ \cite{lpss1}.  This 
Lagrangian can be obtained as a consistent truncation of a Lagrangian 
involving multiple scalars and cosmological terms, in a manner described in 
\cite{lpsol}.  Domain-wall solutions in 4-dimensional supergravities were
studied earlier in \cite{cgr,g,c}. Type II domain walls in $D=10$ were related
to D8-branes in \cite{pw}.  Domain-wall structures in generic dimensions and
their dimensional reduction were studied in \cite{ads}. 

   The above dimensional reduction of the ten-dimensional solution on K3 may 
be implemented also at the level of the supergravity theory itself.  This 
generalised reduction procedure exploits the cohomology class $H^2(M_4,\R)$ 
of the compactifying 4-manifold $M_4$.  Whereas for $T^4$ it has 
dimension 6, corresponding to three self-dual and three anti-self-dual 
harmonic 2-forms, for K3 it has dimension 22.  These comprise 3 self-dual
covariantly-constant 2-forms $\Omega_{2+}^{(i)}$, one of which is the
K\"ahler form, and 19 anti-self-dual harmonic 2-forms $\Omega_{2-}^{(\a)}$,
which are not covariantly constant.  For simplicity, we shall consider the
case where we choose just one of the three covariantly-constant self-dual
2-forms, say $\Omega_{2+}^{(3)}$, for the generalised reduction.  Locally,
we may write $\Omega_{2+}^{(3)}=dw$, and the ans\"atze for the
ten-dimensional potentials are 
\bea
\hat{\cal A}_1^{(1)}(x,z) &=& m\, \omega + {\cal A}^{(1)}(x)\ ,\nn\\
\hat A_3(x,z)&=& A_3(x) + \sum_{i=1}^3 A_1^{(i)}(x)\wedge \Omega_{2+}^{(i)} +
\sum_{\a=1}^{19} A_1^{(\a)}(x)\wedge \Omega_{2-}^{(\a)} \ ,\label{k3ans}\\
\hat A_2^{(1)}(x,z) &=& A_2^{(1)}(x) + \sum_{i=1}^3 A_0^{(1i)}(x)\, 
\Omega_{2+}^{(i)} +\sum_{\a=1}^{19} A_0^{(1\a)}(x)\, \Omega_{2-}^{(\a)} 
\ .\nn
\eea
The ansatz for the metric will be the standard one, with the 58 parameters 
for Ricci-flat metrics on K3 becoming $x$-dependent six-dimensional scalar 
fields.  Note that there are no vector potentials coming from the metric, 
since the first Betti number of K3 is zero.

     The spectrum of massive and massless fields can be determined in the 
same manner as we did previously for the toroidal compactification.  
Substituting the ans\"atze (\ref{k3ans}) into (\ref{csd10}), we see from an 
examination of the terms linear in fields that $A_2^{(1)}$ becomes massive, 
by eating the field $A_1^{(i)}$ with $i=3$, and also the axion $A_0^{(1i)}$
with $i=3$ becomes massive. (The last result follows because 
$\Omega_{2-}^{(\a)}\wedge \Omega_{2+}^{(i)}=0$ and $\Omega_{2+}^{(i)} \wedge 
\Omega_{2+}^{(j)} = 2 \Omega_4\, \delta^{ij}$.)  All the other 
six-dimensional fields will be massless.  There is also a cosmological term 
coming from the kinetic term for the ten-dimensional 2-form field strength.
The theory in $D=6$ has $N=1$ supersymmetry.

    Note that in the K3 compactification that we have been discussing, we
chose one of the covariantly-constant harmonic 2-forms for the generalised 
reduction procedure.  In principle, we could instead choose one of the 
remaining 19 anti-self-dual harmonic 2-forms, which will not be covariantly 
constant.  This will also give rise to a massive supergravity in 6 dimensions.
However, it was evident in our previous discussion of the reduction of the 
intersecting 5-brane solution in $D=10$ that the covariant-constancy of the 
harmonic 2-form played an essential role in the solution.  Thus a 
generalised reduction using an harmonic form that is not covariantly 
constant will give rise to a massive supergravity that does not admit 
domain-wall solutions of the kind we are considering in this paper.  Thus 
here, and in the generalised reductions in subsequent sections, we shall 
concentrate on those associated with covariantly-constant harmonic forms.

\subsection{$T^6$ and Calabi-Yau compactifications}

     The 6-torus has non-vanishing cohomology $H^p(T^6,\R)$ for all $0\le 
p\le 6$.  On the other hand, Calabi-Yau manifolds of real dimension 6 have 
non-vanishing cohomology for $p=0,2,3,4,6$. Thus in either case we may 
consider generalised reductions based on the cohomology classes
$H^2(M_6,\R)$, $H^3(M_6,\R)$ or $H^4(M_6,\R)$.  The procedure for 
implementing the generalised dimensional reduction is similar to the one we 
described in section (2.1), with an additional term of the form $m\omega$ added 
to the relevant ten-dimensional potential, where $d\omega$ represents the 
non-trivial 2'nd, 3'rd or 4'th cohomology.  The spectrum of massive fields 
in the dimensionally-reduced theory is again governed by the structure of 
the $m$-dependent bilinear terms in the four-dimensional Lagrangian.  For 
example, if we make a generalised ansatz for $\hat {\cal A}_1^{(1)}$, we find 
that the field $A_2^{(1)}$ becomes massive, as do certain axionic fields.  
The resulting theory is a massive supergravity in $D=4$, with $N=8$ 
supersymmetry in the case of a toroidal compactification, and $N=2$ in the
Calabi-Yau case. Note that if instead the ansatz for $\hat A_3$ or $\hat
A_2^{(1)}$ is generalised, there will also be bilinear terms in the $D=4$
Lagrangian coming from $d\hat A_3\wedge d\hat A_3 \wedge \hat A_2^{(1)}$ in
$D=10$. These are associated with topological mass terms. 

        Having obtained the massive supergravities in $D=4$ {\it via}
various generalised dimensional procedures, it is interesting to study their
vacuum solutions, namely the domain walls.   Note that unlike the cases in
$D=6$ discussed in the previous subsection, in order to have a domain wall
solution in $D=4$ it is not actually essential to make a generalised
compactification.  This is because a 4-form field strength in $D=4$ is dual
to a cosmological term, and hence can be used to construct an electric
domain wall, which is nothing but the vertical dimensional reduction of the
membrane in $D=10$.  The compactifying space can be any Ricci-flat
6-manifold. However, if we do instead consider the generalised ansatz for
the field strength, the membrane in $D=10$ will not survive the
compactification since the solution is incompatible with the generalised
ansatz, and the domain wall solutions in $D=4$, which are solitonic, arise
as solutions from the cosmological terms. These domain-wall solutions in
$D=4$ massive supergravities have their origins as intersecting solitonic
$p$-branes in $D=10$, as we explained in the examples of $T^4$ or K3
compactifications in section 2.1.  We shall first discuss the $T^6$
compactification, and then show that it can easily be extended to a
Calabi-Yau compactification.  We start with a generalised dimensional
reduction where the 1-form field potential acquires an extra term,
corresponding to $\hat F_2^{(1)} = \Omega_2 + \cdots$, with 
\be
\Omega_2=m_1\, dz_2\wedge dz_3 + m_2\, dz_4\wedge dz_5 + 
m_3\, dz_6\wedge dz_7\ .\label{j}
\ee
In $D=10$, there exists a solution of three intersecting 6-branes, with the 
2-form field strength given by (\ref{j}).  The metric and the dilaton 
$\phi_1$ of the solution are given by
\bea
ds_{10}^2 &=& (H_1H_2H_3)^{-\ft18} dx^\mu dx^\nu \eta_{\mu\nu} +
(H_1H_2H_3)^{\ft78} dy^2 + H_1^{\ft78} (H_2H_3)^{-\ft18} (dz_2^2 + dz_3^2)
\nonumber\\
&& + H_2^{\ft78} (H_1H_3)^{-\ft18} (dz_4^2 + dz_5^2) +
H_3^{\ft78} (H_1H_2)^{-\ft18} (dz_6^2 + dz_7^2)\ ,\label{3sixbranes}\\
e^{\phi_1} &=& (H_1H_2H_3)^{\ft34}\nn
\eea
where $H_\a = 1 + m_\a |y|$. This solution can be compactified on the
6-torus, parameterised by $z^i$ for $2\le i \le 7$, giving rise to a
domain-wall solution in $D=4$: 
\bea
ds_4^2 &=& (H_1 H_2 H_3)^{\ft12} + (H_1 H_2 H_3)^{\ft32} dy^2\ ,\nonumber\\
e^{-\ft12 \varphi_\a} &=& H_\a (H_1 H_2 H_3)^{\ft34}\ ,\label{d4dw}
\eea
where $\varphi_\a=\vec c_a\cdot \vec \phi$, with $\vec c_1=\vec b_{123}$,
$\vec c_2 = \vec b_{145}$ and $\vec c_3 = \vec b_{167}$. The domain wall
(\ref{d4dw}) arises as a solution of the 4-dimensional Lagrangian
\be 
e^{-1} {\cal L} = R - \ft12 (\del \vec \phi)^2 -\ft12 
\sum_{\a=1}^{3} m_\a^2 e^{\vec c_\a \cdot \vec \phi}\label{d4lag}
\ee 
which can be obtained from the generalised dimensional reduction of the theory.

     In order to extend the above discussion to the generalised Calabi-Yau
compactification, we can consider the special case where the parameters
$m_\a$ are all equal.  In this case, in the solution of the three
intersecting 6-branes in $D=10$, the 2-form field strength is taken to be
proportional to the K\"ahler form $J$ and the metric becomes 
\be
ds_{10}^2 = H^{-\ft38} dx^\mu dx^\nu \eta_{\mu\nu} + H^{\ft{21}{8}}dy^2 +
H^{\ft58} ds_6^2\ ,
\ee
where $ds_6^2$ is the metric of the Calabi-Yau manifold.  As in the 
compactification on the 4-torus that we discussed previously, the form of 
the three-term expression (\ref{j}) for the 2-form field strength is 
uniquely singled out, modulo coordinate relabellings, by the requirement that 
it should give rise to an isotropic form for the six-dimensional internal 
metric when the charges are set equal.  Thus compactifying
this solution on the Calabi-Yau manifold gives rise to a single-scalar 
domain-wall solution in $D=4$ with $\Delta = 4/3$, which is the vacuum
solution of the 4-dimensional $N=2$ massive supergravity theory. 

      The analysis of domain-wall solutions of other massive supergravities
in $D=4$, coming from the dimensional reductions with the 2-form or the 3-form
potential taking the extra generalised ansatz, is analogous.   In the
case of a toroidal compactification, the extra term for the 2-form potential
corresponds to an harmonic 3-form that may be taken to be
\bea 
\Omega_3&=& m_1\, dz_2\wedge dz_3 \wedge dz_4 -
m_2 \, dz_2\wedge dz_6 \wedge dz_7 +
m_3\, dz_3\wedge dz_5 \wedge dz_7 \nonumber\\
&&- m_4\, dz_4\wedge dz_5 \wedge dz_6\ .\label{l}
\eea
We find that there is a solution of four intersecting 5-branes in $D=10$ if
the 3-form field strength is set equal to $\Omega_3$:
\bea
ds^2_{10}&=&(H_1H_2H_3H_4)^{-\ft14}\Big( dx^\mu dx^\nu \eta_{\mu\nu} +
 (H_1H_2H_3H_4) dy^2 + (H_1H_2) dz_2^2
\nonumber\\
&&\qquad + (H_1H_3)dz_3^2+ (H_1H_4)^{\ft34}dz_4^2 + 
(H_3H_4)^{\ft34} dz_5^2 \label{4fivebranes}\\
&&\qquad + (H_2H_4)dz_6^2 + (H_2H_3)dz_7^2\Big)\nonumber\\ 
e^{\phi_1}&=&(H_1H_2H_3H_4)^{-\ft12}\nonumber
\eea 
Compactification of the solution on the 6-torus gives rise to a domain wall
in $D=4$, which is a solution for the 4-dimensional Lagrangian
(\ref{d4lag}) (but with $\a$ now running over $1\,\ldots,4$, and $\vec
c_1=\vec a_{1234}$, $\vec c_2 = \vec a_{1267}$, $\vec c_3 = \vec a_{1357}$,
$\vec c_4 = \vec a_{1456}$). When the $m_\a$ are all equal, the discussion
can be extended to a generalised compactification on a Calabi-Yau manifold.
Again, the requirement that the isotropic form for the internal metric 
should arise in the 
equal-charge limit demands that the 3-form $\Omega_3$ have the four-term 
structure given in (\ref{l}).
In this case, the metric (\ref{4fivebranes}) for the solution of the four
intersecting 5-branes becomes 
\be
ds_{10}^2 = H^{-1} dx^\mu dx^\nu \eta_{\mu\nu} + H^{3} dy^2 +
         H\, ds_6^2\ ,\label{solextra}
\ee
where $ds_6^2$ is the metric for the Calabi-Yau manifold.  The 3-form 
$\Omega_3$ given by (\ref{l}) on the 6-torus will now be replaced by $m 
\Lambda$, where $\Lambda$ is the real part of the complex holomorphic 3-form
\be
\ft16 \epsilon^{abc}\, d\zeta_a\wedge d\zeta_b \wedge d\zeta_c\ ,
\label{complex}
\ee
where the three complex coordinates $\zeta_a$ can be related to the six 
real coordinates $z_i$ by $\zeta_1=z_2 + {\rm i}\, z_5$, $\zeta_2=z_3 + {\rm
i}\, z_6$ and $\zeta_3=z_4 + {\rm i}\, z_7$.  Thus the structure of this
holomorphic 3-form on the Calabi-Yau space coincides with the structure that
we had to choose for the harmonic 3-form (\ref{l}) on the 6-torus in order
to be able to achieve an isotropic limit for the internal metric on the
torus.  This is a rather striking indication that the type IIA string
exhibits special features that conspire with the properties of Calabi-Yau
spaces to make the construction of the associated massive supergravity
possible.  Similar remarks apply to the other Calabi-Yau and K3
compactifications that we discussed previously.  The ten-dimensional
solution (\ref{solextra}) will reduce to a single-scalar domain-wall
solution with $\Delta=1$ under the compactification that we are discussing
here. 

       The generalised ansatz for the 3-form potential in a toroidal 
compactification corresponds to adding an harmonic 4-form to $F_4$ that 
can be taken to have the form
\be
\Omega_4= m_1\, dz_2 \wedge dz_3 \wedge dz_4 \wedge dz_5 +
   m_2\, dz_2 \wedge dz_3 \wedge dz_6 \wedge dz_7 +
   m_3\, dz_4 \wedge dz_5 \wedge dz_6 \wedge dz_7\ .\label{k}
\ee
The solution that survives the compactification has the 4-form
field strength equal to $\Omega_4$.  In this case, it describes three
intersecting 4-branes: 
\bea
ds_{10}^2 &=& (H_1 H_2 H_3)^{-\ft38} dx^\mu dx^\nu \eta_{\mu\nu} +
(H_1 H_2 H_3)^{\ft58} dy^2+ (H_1H_2)^{\ft58} H_3^{-\ft38} (dz_2^2 + dz_3^2)
\nonumber\\
&& + (H_1H_3)^{\ft58} H_2^{-\ft38} (dz_4^2 + dz_5^2) +
+ (H_2H_3)^{\ft58} H_1^{-\ft38} (dz_6^2 + dz_7^2) 
\ ,\label{3fourbranes}\\
e^{\phi_1}&=& (H_1H_2H_3)^{\ft14}\ .\nonumber
\eea
When the $m_\a$'s are all equal, we can extend the result to a Calabi-Yau 
compactification.   The 10-dimensional solution (\ref{3fourbranes}), whose
metric now becomes 
\be
ds_{10}^2 = H^{-9/8} dx^\mu dx^\nu\eta_{\mu\nu} + H^{15/8}
dy^2 + H^{7/8}ds_6^2\ ,
\ee
which can be compactified on a Calabi-Yau manifold with the Ricci-flat metric
$ds_6^2$, giving rise to a single-scalar domain-wall solution with
$\Delta=4/3$ in the 4-dimensional $N=1$ massive supergravity.  The 4-form
$\Omega_4$, which is given by (\ref{k}) on the 6-torus, will be replaced by $m
J\wedge J$, where $J$ is the K\"ahler form on the Calabi-Yau space.  Again, 
the structure of the 4-form (\ref{k}) on the 6-torus, required in order that 
there exist an equal-charge limit with an isotropic metric on the 6-torus,
is precisely what is needed in order to allow a generalisation to a 
Calabi-Yau compactification.

\section{Generalised reduction of M-theory and the type IIB string}

    M-theory has only a 4-form field strength, and so a generalised
dimensional reduction requires a compactification on a manifold with a
non-vanishing 4'th cohomology class.  The simplest generalised reduction is
obtained by compactifying M-theory on a 4-manifold $M_4$, with the 3-form
potential having an additional term whose exterior derivative is
proportional to the 4-volume form on $M_4$.  The case of a $T^4$
compactification was discussed in \cite{clpst}, and gives rise to a
$D=7$, $N=2$ massive supergravity, whose vacuum solution, the domain wall,
is the vertical dimensional reduction of the 5-brane in $D=11$. In
\cite{lpdomain}, the discussion was extended to a generalised K3
compactification, giving rise to $N=1$ massive supergravity in $D=7$ with a
single topological mass term for the 4-form field strength in the
supergravity multiplet.  This $N=1$ massive supergravity in $D=7$ was
constructed earlier in \cite{mtn}, with an additional gauge parameter.  The
supersymmetric vacuum solution with both the gauge and mass parameters was
constructed in \cite{lpss2}. 

      For compactifications to $D=4$, the possibilities for the internal
7-manifold include $T^7$, Y$\times S^1$, K3$\times T^3$, and Joyce manifolds
J$_7$, where Y denotes a six-dimensional Calabi-Yau manifold. Note that the
generalised dimensional reductions on these manifolds have properties
essentially similar to those demonstrated in the previous section.  We shall
therefore not discuss all the manifolds in detail, but instead we shall
focus on the simplest and most illuminating example, namely the case where
M-theory is compactified on a Joyce manifold.  Seven-dimensional Joyce
manifolds are simply-connected compact manifolds with $G_2$ holonomy that
admit Ricci-flat metrics \cite{j}.  It follows from the decomposition of the
tangent-space group $SO(7)$ under the $G_2$ subgroup that the 8-dimensional
spinor representation decomposes into 7 + 1 under the holonomy group, and
therefore there is one covariantly-constant Majorana spinor, $\eta$.  This
implies that the compactified four-dimensional supergravity theory will have
$N=1$ supersymmetry.  The cohomology structure of the manifold is specified
by the 2'nd and 3'rd Betti numbers, with $b_0=1, b_1=0, b_4=b_3, b_5=b_2,
b_6=0, b_7=1$.  Included in the 3'rd cohomology is a covariantly-constant
3-form which can be constructed using $\eta$, as $\Omega_{ijk} = \bar\eta
\Gamma_{ijk} \eta$.  Its dual, $\Omega_4= *\Omega_3$, is also covariantly
constant, and this is the form that we shall use for the generalised
reduction of M-theory. 

     The bosonic Lagrangian for the low-energy limit of M-theory is
\be
{\cal L} = \hat e \hat R -\ft1{48} \hat e\, \hat F_4^2 + \ft16 \hat F_4 
\wedge \hat F_4 \wedge \hat A_3\ ,\label{mtheory}
\ee
where $F_4=dA_3$.  The generalised ansatz for the dimensional reduction of 
$\hat A_3$ is
\be
\hat A_3(x,z) =  m\, \omega + A_3(x) + \sum_{\a=1}^{b_2} A_1^{\a}(x)\wedge
\Omega_2^{\a} + \sum_{i=1}^{b_3} A_0^{(i)}(x) \, \Omega_3^{i}\ ,
\label{jans}
\ee
where $d\omega=\Omega_4$. Note that the dual of $\Omega_4$ is one of the 
3-forms $\Omega_3^{i}$ included in the expansion.  For convenience, we may
assume that it is $\Omega_3^1$.  We shall choose a basis for the
$\Omega_3^i$ such that $\int \Omega_3^i \wedge *\Omega_3^j = \delta^{ij}$.  
In addition to the 3-form $A_3$, the $b_2$ 1-form potentials $A_1^{\a}$ and
$b_3$ axions $A_0^{i}$ in $D=4$, there will be $b_L+1$ scalars, coming from 
the $b_L$ transverse traceless zero modes of the Lichnerowicz operator on 
the Joyce manifold J$_7$ together with the conformal scaling mode.  These
$b_L+1$ zero modes correspond to the independent Ricci-flat deformations of
$J$. It was shown in \cite{gpp} that $b_L$ is equal to $b_3-1$.   Since the 
4-form field strength $\hat F_4$ in $D=11$ has no Chern-Simons corrections, 
the only source of mass terms in the $D=4$ theory obtained by this generalised 
dimensional reduction is the $\hat F_4\wedge \hat F_4 \wedge \hat A_3$ term 
in $D=11$.  In view of the fact that $\int \Omega_4\wedge 
\Omega_3^i=\delta^{1i}$, we find that that it gives a contribution ${\cal 
L}_{\rm mass} = \ft1{3856}m A_0^{(1)} \epsilon^{\mu\nu\rho\sigma} 
F_{\mu\nu\rho\sigma}$ to the four-dimensional Lagrangian.  This term implies 
that the axion $A_0^{(1)}$ is massive.  To see this, we can look at the 
relevant quadratic terms in the four-dimensional Lagrangian:
\be
{\cal L} \sim -\ft12 \sum_{i=1}^{b_3} (\del\phi_i)^2 
-\ft12 \sum_{i=1}^{b_3} (\del A_0^{(i)})^2 
-\ft1{48} F_4^2 -\ft1{12} \sum_{\a=1}^{b_2} (F_2^{(\a)})^2 +
\ft1{3856}m A_0^{(1)} \epsilon^{\mu\nu\rho\sigma} F_{\mu\nu\rho\sigma}
\ee
The equation of motion for $A_3$ implies that $F_{\mu\nu\rho\sigma} =
\ft1{144} m A_0^{(1)} \epsilon_{\mu\nu\rho\sigma}$, and hence the Lagrangian 
for $A_0^{(1)}$ has the form $-\ft12(\del A_0^{(1)})^2 -\ft1{(12)^4} m^2 
(A_0^{(1)})^2$.  

   In order to obtain the domain-wall solution in $N=1$ massive supergravity 
in $D=4$ that we constructed above, and to relate this solution to a solution 
in $D=11$ dimensions, we first consider a generalised compactification on a
7-torus.  In order to be able to generalise the compactification to a Joyce 
manifold, it is necessary to introduce a harmonic 
4-form on the 7-torus that has seven independent terms, of the form
\bea
\Omega_4 &=& m_1\, dz_1\wedge dz_2 \wedge dz_3 \wedge dz_7 +
m_2\, dz_1\wedge dz_2 \wedge dz_4 \wedge dz_5\nonumber\\
&&+m_3\, dz_1\wedge dz_3 \wedge dz_4 \wedge dz_6 +
m_4\, dz_1\wedge dz_5 \wedge dz_6 \wedge dz_7\nonumber\\
&&+m_5\, dz_2\wedge dz_3 \wedge dz_5 \wedge dz_6 +
m_6\, dz_2\wedge dz_4 \wedge dz_6 \wedge dz_7\label{d11omega4}\\
&&+m_7\, dz_3\wedge dz_4 \wedge dz_5 \wedge dz_7\ .\nonumber
\eea
As in the previous examples in the type IIA string, the structure 
(\ref{d11omega4}) is dictated uniquely, up to coordinate redefinitions, by 
the requirement that the associated solution have a limit where the metric 
on the 7-torus have an isotropic form when the charges are set equal.
The solution in $D=11$, with the 4-form field strength $\hat F_4$
equal to $\Omega_4$, describes seven intersecting fivebranes: 
\bea
ds_{11}^2 &=& (H_1\cdots H_7)^{-\ft13}\Big(dx^\mu dx^\nu\eta_{\mu\nu} +
(H_1\cdots H_7) dy^2 + (H_1H_2H_3H_4) dz_1^2 \nonumber\\
&& (H_1H_2H_5H_6) dz_2^2 + (H_1 H_3 H_5H_6) dz_3^2 +
(H_2H_3H_6H_7) dz_4^2\label{7fivebranes}\\
&& (H_2H_4H_5H_7) dz_5^2 + (H_3H_4H_5H_6) dz_6^2 +
   (H_1H_4H_6H_7) dz_7^2\Big)\ ,\nonumber
\eea
where $H_\a= 1 + m_\a |y|$.   Thus compactifying this solution on the
7-torus parameterised by $z^i$ gives rise to a domain-wall solution in
$D=4$: 
\bea
ds_4^2 &=& (H_1\cdots H_7)^{\ft12} dx^\mu dx^\nu\eta_{\mu\nu} +
       (H_1\cdots H_7)^{\ft32} dy^2\ ,\nonumber\\
e^{-\ft12 \vec c_\a \cdot \vec\phi} &=& H_\a (H_1\cdots H_7)^{\ft43}\ ,
\eea
where 
\bea
&&\vec c_1 = \vec a_{1237},\qquad \vec c_2 = \vec a_{1245}\ ,\qquad
\vec c_3 = \vec a_{1346},\qquad \vec c_4 = \vec a_{1567}\ ,\nonumber\\
&&\vec c_5 = \vec a_{2356},\qquad \vec c_6 = \vec a_{2467}\ ,\qquad
\vec c_7 = \vec a_{3457}.
\eea
Supersymmetric $p$-brane solutions with seven charges in $D=4$, $N=8$ 
supergravity were first found in \cite{lpsol}.   When all the charge
parameters $m_\a$ are equal, the metric on the 7-dimensional internal space
becomes proportional to $dz_i\, dz_i$, which can be replaced by an arbitrary
Ricci-flat metric, provided that it admits a covariantly-constant 4-form
that can replace the expression (\ref{d11omega4}) on the 7-torus.  
Remarkably, any Joyce manifold has precisely such a 4-form.  We may first
consider the 3-form dual to this, which is given, as we remarked earlier, by
$\Omega_{ijk}=\bar \eta \Gamma_{ijk} \eta$.  As is well known \cite{ct,ers},
the quantities $\Omega_{ijk}$ (in a tangent-space frame) generate the
multiplication table of the imaginary octonions $o_i$, namely $o_i\, o_j =
-\delta_{ij} + \Omega_{ijk} \, o_k$.  In a suitable basis, the non-zero
components of $\Omega_{ijk}$ are specified by 
\be 
\Omega_{147}=\Omega_{257}=\Omega_{367}=\Omega_{126}=\Omega_{234}=
\Omega_{315}=\Omega_{456}=1\ ,\label{octonion}
\ee
which, after dualising, is easily seen to give the same structure as  
(\ref{d11omega4}).  The 11-dimensional metric for the seven intersecting
5-branes then becomes, after setting the $m_\a$'s equal,
\be 
ds_{10}^2 = H^{-\ft73} dx^\mu dx^\nu\eta_{\mu\nu} + H^{\ft{14}3} dy^2
+ H^{\ft53} ds_7^2\ ,
\ee
where $ds_7^2$ is the Ricci-flat metric on the Joyce manifold.  This 
solution reduces to a single-scalar domain wall with $\Delta=4/7$ upon 
compactification to $D=4$.  Thus we see also in this case that M-theory has 
special features that are precisely what is needed in order to allow the 
generalised compactification on a 7-dimensional Joyce manifold.

     We shall now consider generalised compactifications of the
10-dimensional type IIB theory, which, it has been argued, should be
accorded a similar fundamental status to that of M-theory \cite{v}. The
bosonic fields of type IIB supergravity consist of a self-dual 5-form, two
3-form and a 1-form field strength, together with the metric and dilaton.  
A generalised compactification using the 1-form field strength $F_1=d\chi$ on
a circle, with the ansatz $\chi(x,z) = m\,z + \chi(x)$, was first considered in
\cite{bdgpt}. It gives rise to an $N=2$ massive supergravity in $D=9$, which is
T-dual to the $S^1$ compactification \cite{bdgpt} of 10-dimensional $N=2$ 
massive supergravity \cite{r}.  In this procedure, the axion $\chi(x)$ is
eaten by the Kaluza-Klein vector, which becomes massive.  In a
compactification on an $n$-torus, the generic ansatz $\chi(x,z)= m_i \, z_i
+ \chi(x)$, where $i$ runs over the dimension of the compactified space,
again generates only one cosmological term, since in the first step of
reduction where a mass parameter $m_i$ is non-zero, the $\chi$ field is
again eaten, and thus cannot generate any further cosmological terms in the
subsequent reduction steps. (Put another way, the reduction gives
``cosmological terms'' of the form $-\ft12 m_r^2\, e^{\vec c_r\cdot\vec\phi}
-\ft12 \sum_{i>r} (m_i+\chi_i)^2\, e^{\vec c_i\cdot \vec\phi}$, where $m_r$
is the first non-vanishing mass parameter, and $\chi_i$ are axions coming
from the dimensional reduction of Kaluza-Klein vectors.  Constant shift
transformations of these axions enable the constants $m_i$ for $i>r$ to be
absorbed, leaving only the cosmological term for $m_r$.)  The spectrum of
massive fields in $10-n$ dimensions does depend, however, on the value of
$r$ labelling the first non-zero component $m_r$ of the parameters $m_i$ 
\cite{clpst}.  This is because each subsequent reduction step after the
appearance of the massive vector potential will generate an additional
massive scalar from it, and so the number of these additional massive
scalars depends on the step at which the massive vector appears.  In other
words, ordinary Kaluza-Klein reduction and generalised Kaluza-Klein
reduction do not commute in general, which was observed in the case of $S^1$
compactification in \cite{clpst}. 

     Generalised dimensional reductions using either of the 3-form field
strengths in the type IIB theory are analogous to those that we presented
for the type IIA theory. We shall not consider such a compactification on
Calabi-Yau spaces further; instead, we shall consider the example of a
generalised compactification to three dimensions on a 7-dimensional Joyce
manifold, using the covariantly-constant 3-form $\Omega_3$ that we discussed
previously. The NS-NS 3-form $F_3$ and the R-R 3-form $\widetilde F_3$ are
given in terms of potentials by 
\be
F_3=dA_2\ ,\qquad \widetilde F_3 = d\widetilde A_2 + \chi\, dA_2\ .
\label{2bcs}
\ee
The trilinear term in the type IIB theory is of the form $A_4\wedge d A_2 
\wedge d\widetilde A_2$.  This term will not give rise to any mass 
contributions, regardless of whether the generalised reduction is applied to 
$A_2$ or $\widetilde A_2$, since, after substituting in the Kaluza-Klein 
expansions it will give bilinear terms proportional to the 
triple-intersection numbers $\int_J \Omega_2^i\wedge \Omega_2^j \wedge 
\Omega_3$.  Since none of the harmonic 2-forms $\Omega_2^i$ are covariantly 
constant, it follows that none of the harmonic 4-forms $\Omega_2^i\wedge
\Omega_2^j$ are covariantly constantant, and therefore they will have zero 
intersection with $\Omega_3$.  The only source of a mass term is therefore 
from the Chern-Simons correction in (\ref{2bcs}).  Thus the axion $\chi$ 
will acquire a mass if the generalised reduction is applied to the NS-NS 
potential $A_2$, whilst it will remain massless if instead the R-R potential 
$\widetilde A_2$ is used.  In either case, there will be a cosmological term 
in the 3-dimensional $N=2$ supergravity theory.  The domain wall solutions 
of these two theories can both be oxidised back to $D=10$, becoming seven 
intersecting 5-branes in the type IIB theory.  These carry NS-NS charges in 
the former case, and R-R charges in the latter.

     It now remains to discuss the generalised ansatz for the self-dual
5-form, which is associated with a self-dual 3-brane in $D=10$.  In fact the
self-dual 3-brane can be viewed as being analogous to the M-branes of
$D=11$, in that it is not coupled to the dilaton. The self-dual 3-brane
solution, with its charge uniformly distributed on a 5-dimensional
hyperplane in the transverse space, is given by 
\bea 
ds_{10}^2 &=& H^{-\ft12} dx^\mu dx^\nu\eta_{\mu\nu} + H^{\ft12} (dy^2 + 
ds_5^2)\ ,\nonumber\\
\hat F_5 &=& m (H^{-2}\, \epsilon_4\wedge dy + \Omega_5)\ ,
\eea
where $\Omega_5$ is the volume form for the 5-dimensional 
metric $ds_5^2$, and $\epsilon_4$ is the volume form for the 3-brane world
volume.  As usual, the harmonic function $H$ takes the form $H=1+m |y|$. The
flat metric $ds_5^2=dz_i\, dz_i$ on the hyperplane can be replaced by any
Ricci-flat metric on a compactifying 5-dimensional space, giving rise to a
domain wall in $D=5$.  At the level of the theory itself, the
compactification to a supergravity in $D=5$ that admits the domain wall
solution requires a generalised ansatz of the form
\be
\hat A_4(x,z) = m\, \omega_4 + \cdots \ ,
\ee
where $d\omega_4 = H^{-2}\, \epsilon_4\wedge dy+\Omega_5$, and the dots
represent the standard harmonic expansions for the lower-dimensional
massless fields.  As far as we are aware, there are no particularly
noteworthy 5-dimensional Ricci-flat compact spaces. 

\section{Discussion and conclusions}

     In this paper, we have studied the generalisations of the usual 
Kaluza-Klein ansatz that are necessary in order to generate
lower-dimensional massive theories that admit domain wall solutions.  The
new characteristic of the generalised ansatz is that there are additional terms
in the gauge potential, whose exterior derivatives are constant multiples
of certain harmonic forms on the compactified space.  As long as the gauge
potential enters the higher-dimensional equations of motion everywhere 
{\it via} its
derivative, this generalised dimensional reduction will be consistent, as in
the case of standard Kaluza-Klein procedure. We focussed our attention on
several such generalised reductions, for M-theory and for the type IIA and
type IIB strings, compactified on certain Ricci-flat manifolds.   In
general, this generalised dimensional reduction procedure will give rise to
a theory in the lower dimensions that contain no $p$-brane solutions with
$p\le (D-3)$.  In the cases we considered, the lower-dimensional theories
admit supersymmetric domain-wall solutions.  Owing to the consistency of the
dimensional reduction procedure, these domain-wall solutions can be oxidised
back to $D=10$ or $D=11$, where they become higher-dimensional $p$-branes,
or intersecting $p$-branes.   A particular example worth mentioning is the
generalised dimensional reduction of M-theory on a 7-dimensional Joyce
manifold J$_7$, where the 4-form field strength has an additional term that is
proportional a covariantly-constant harmonic form on J.  The associated
domain-wall solution in $D=4$ describes seven intersecting 5-branes upon
oxidation to $D=11$, with all seven charges equal.  More general solutions
exist in $D=11$, where the seven charges are independent parameters.  These
solutions can be reduced to $D=4$ on a 7-torus, but, as can be seen from 
(\ref{7fivebranes}), with a metric on the 7-torus in which the seven 
coordinates enter asymmetrically.  They only enter in a symmetrical way, 
allowing the possibility of replacing the 7-torus metric $dz_i\, dz_i$ by 
the Ricci-flat metric on the Joyce manifold, when the charges are equal. In
fact the only way of achieving a solution for intersecting 5-branes in which 
the seven coordinates enter symmetrically is by using exactly seven terms 
in the 4-form field strength, as in (\ref{d11omega4}), with the seven
charges set equal.  Thus we see that special features of $D=11$ supergravity 
are ideally adapted for exploiting the exceptional properties of 
seven-dimensional Joyce manifolds.

      A further issue that arises in connection with the generalised
dimensional reduction discussed in this paper concerns duality.   In
general, the dualities that have been established in the standard
Kaluza-Klein reduction cease to hold in the generalised reduction schemes.
For example, the standard K3 compactification of M-theory is conjectured 
\cite{w} to be related by S duality to the $T^3$ compactification of the
heterotic string.  This can easily be seen to be consistent at the
supergravity level with the counting of the massless fields in $D=7$
corresponding to the two compactifications. In fact the two $D=7$
supergravities are related to each other by local field redefinitions,
involving in particular a sign reversal of the dilaton.  However, this
duality seems break down if we introduce the generalised compactification
procedure.  To see this, note that if the 4-form field strength of $D=11$
supergravity acquires an additional term that is a constant multiple of the
volume form on K3, it gives rise to a 7-dimensional supergravity with a
topological mass term \cite{lpdomain}.  On the other hand, such a
topological mass term cannot arise from the $T^3$ compactification of the
heterotic theory, whether with the standard or the generalised reduction. 
Instead, the generalised ansatz for toroidal compactification of the
heterotic supergravity will give masses to Kaluza-Klein vectors, which does
not happen in the generalised K3 compactification of M-theory.  Similarly we
find that in most other cases too, the dualities that are manifest in
standard Kaluza-Klein reductions break down.  One outstanding problem is
that the massive type IIA supergravity in $D=10$ does not seem to be related
directly to M-theory compactified in any lower dimension \cite{clpst},
although it was shown that its ordinary reduction on $S^1$ is T dual to the
generalised reduction of massless type IIB supergravity on $S^1$
\cite{bdgpt}. This leads to the conjecture in \cite{clpst} that there may
exist a 13-dimensional hypothetical H-theory. 

     However, by contrast there are some examples where duality does seem to
survive the passage to a generalised dimensional compactification. It is
conjectured that the type IIA string compactified on K3 is dual to M-theory
compactified on K3$\times S^1$, using standard Kaluza-Klein reduction.  At
the level of supergravity, this is rather trivial, since it merely states
that the K3 and the $S^1$ compactifications of 11-dimensional supergravity
commute. It was shown in \cite{lpdomain} that this commutative property is
preserved even in the case of generalised dimensional reduction, where the
4-form has an additional term proportional to the K3 volume form.  This
commutativity is rather non-trivial for the generalised dimensional
reduction procedure.  For example, it would no longer hold if we replaced
the K3 compactifying manifold by $T^4$ \cite{lpdomain}. In other words,
M-theory compactified on $T^4\times S^1$ seems to be inequivalent to the
type IIA string compactified on $T^4$, in the case where a generalised
Kaluza-Klein ansatz is used. 

     Another example is provided by the generalised reduction of M-theory
compactified on J$_7\times S^1$.  As we saw in section 3, the generalised
compactification of M-theory on a 7-dimensional Joyce manifold J$_7$ gives
one massive scalar field.  It is easy to show that the generalised
compactification of M-theory on J$_7$ commutes with the standard
Kaluza-Klein compactification on $S^1$.  In other words, as in the K3 case,
the generalised compactication of M-theory on J$^7\times S^1$ is dual to
that of type IIA compactified on J$_7$, where it is the 4-form field
strength ans\"atze that are generalised. It was observed that the field
content of the standard Kaluza-Klein compactification of the type IIA and
type IIB theories compactified on J$_7$ are the same, which leads to the
conjecture that the J$_7$ compactification of the two theories are
equivalent \cite{pt2}.  We note that since the 3-form and 4-form cohomology
classes are Hodge dual on J$_7$, the duality conjectured in \cite{pt2} may
be extended to the generalised compactification.  In the type IIB theory
there are two 3-form field strengths, one of which is NS-NS and the other
R-R. When the NS-NS 3-form acquires an additional term that is proportional
to the covariantly constant 3-form of the J$_7$ manifold, the resulting
3-dimensional $N=2$ massive supergravity contains a massive scalar, which
was the axionic scalar in type IIB in $D=10$.\footnote{Note that if we
choose instead the R-R 3-form field strength to carry the extra term, the
3-dimensional theory will be $N=2$ massless supergravity, but with a
cosmological term, as discussed in section 3.}  The field content of this
$N=2$ massive supergravity in $D=3$ is identical to that from the
generalised compactification of type IIA on J$_7$, or of M-theory on
J$_7\times S^1$.  Thus we expect that the duality of type IIA and type IIB
conjectured in \cite{pt2} in the context of a standard J$_7$ compactifcation
will hold even in this generalised J$_7$ compactification.  Unlike the
T-duality of the type IIA and type IIB theories compactified on a circle,
which is perturbative in all string orders, the duality of the two theories
compactified on J$_7$ is non-perturbative.  This can be seen easily in the
context of the generalised compactification.  The domain-wall solution of
the type IIA theory compactified on J$_7$ comes from the dimensional
reduction of seven intersecting R-R 4-branes in $D=10$, whilst the
domain-wall solution of the type IIB theory compactified on J$_7$ comes from
the dimensional reduction of seven NS-NS 5-branes in $D=10$.  The duality of
the theories implies that the seven intersecting R-R type IIA 4-branes are
dual to the seven intersecting NS-NS type IIB 5-branes.  Dualities between
NS-NS and R-R $p$-branes are characteristically non-perturbative. 

     Finally, we turn to a discussion of the supersymmetry of the
domain-wall solutions in the massive supergravities that arise from the
generalised dimensional reduction of M-theory or the type II theories. As we
mentioned earlier, these domain-wall solutions can be oxidised back to
higher dimensions, where they become intersecting $p$-branes. The fraction
of supersymmetry that is preserved can then be directly calculated from the
higher-dimensional supersymmetry transformation rules.  This fraction of
preserved supersymmetry will remain unchanged under toroidal dimensional
reduction, even in the case of the generalised Kaluza-Klein ansatz that we
are considering here.  In fact the theory itself retains maximal
supersymmetry under this reduction. On the other hand, for compactifications
involving K3, Calabi-Yau, or Joyce manifolds, the lower-dimensional theories
will have $\ft12$, $\ft14$ or $\ft18$ of the maximal supersymmetry
respectively.  The fraction of the lower-dimensional supersymmetry that the
domain-wall solution preserves depends on the number of cosmological terms
used in its construction.  Although the domain-wall solution may become a
multi-charge intersecting $p$-brane solution in the higher dimension, it is
nevertheless a single-charge solution in the lower dimensional theory
obtained by K3, Calabi-Yau or Joyce manifold compactification, and hence it
preserves half of the lower-dimensional supersymmetry.

\end{document}